\documentclass[twocolumn,superscriptaddress,showpacs,nofootinbib,preprintnumbers,secnumarabic,amssymb, nobibnotes, aps, prd]{revtex4-2}
\usepackage[utf8]{inputenc}
\usepackage{graphicx}
\usepackage{latexsym,amsmath,amssymb,amsthm,lmodern,float,url}
\usepackage{bbm}
\usepackage{natbib}
\usepackage{color}
\usepackage{qcircuit}
\usepackage{microtype}
\usepackage{import}
\usepackage{bbold}
\usepackage[plain]{fancyref}
\usepackage{varioref}
\usepackage{slashed}
\usepackage{multirow}
\usepackage{tikz}
\usepackage{scrextend}
\usepackage{braket}

\usepackage{algorithm}
\usepackage{algpseudocode}
\usepackage{algorithmicx}

\usetikzlibrary{shapes}
\usetikzlibrary{positioning}
\usepackage[normalem]{ulem}

\usepackage{qcircuit}

\newcommand{\fig}[1]{Fig.~\ref{fig:#1}}

\usepackage[colorlinks=true,backref=false, linktocpage=true,
citecolor=blue,urlcolor=blue,linkcolor=blue,pdfpagemode=UseOutlines]{hyperref}
\hypersetup{%
  bookmarksnumbered=true,
  pdftitle = {},
  pdfsubject = {},
  pdfauthor = {},
  pdfkeywords = {}
}

\usepackage{graphicx}% Include figure files
\usepackage{dcolumn}% Align table columns on decimal point
\usepackage{bm}% bold math
\begin{document}

\preprint{FERMILAB-PUB-22-838-T}

\title{Smearing lattice gauge fields on a quantum computer}% Force line breaks with \\

\author{Erik Gustafson}
\affiliation{%
 Fermi National Accelerator Laboratory, Batavia, Illinois 60510, USA
}%

\date{\today}% It is always \today, today,
\begin{abstract}
    Smearing of gauge-field configurations in lattice field theory improves the results of lattice simulations by suppressing high energy modes from correlation functions. In quantum simulations, high kinetic energy eigenstates are introduced when the time evolution operator is approximated such as Trotterization. While improved Trotter product formulae exist to reduce the errors, they have diminishing accuracy returns with respect to resource costs. Therefore having an algorithm that has fewer resources than an improved Trotter formula is desirable. In this work I develop a representation agnostic method for quantum smearing and show that it reduces the coupling to high energy modes in the discrete nonabelian gauge theory $\mathbb{D}_4$.
\end{abstract}
% \begin{abstract}
% An article usually includes an abstract, a concise summary of the work
% covered at length in the main body of the article. 
% \begin{description}
% Secondary publications and information retrieval purposes.
% You may use the \texttt{description} environment to structure your abstract;
% use the optional argument of the \verb+\item+ command to give the category of each item. 
% \end{description}
% \end{abstract}

\maketitle

%\tableofcontents

\textit{1 Introduction.} Classical lattice gauge theory (LGT) offers the ability for high precision determinations of observables such as decay constants, hadron masses, scattering amplitudes below multiparticle thresholds, finite temperature QCD, and equations of state \cite{Detmold:2019ghl,Aoyama:2020ynm,FlavourLatticeAveragingGroupFLAG:2021npn,USQCD:2022mmc,Bazavov:2019lgz,Boyle:2022uba,Kronfeld:2019nfb,Davoudi:2022bnl,Brice_o_2018}.
However, these simulations struggle to extract multiparticle threshold final states and dynamical quantities such as viscosities which encounter sign problems using stochastic methods \cite{deForcrand:2009zkb,Tripolt:2018xeo}.
Quantum computers offer a method to circumvent this sign problem entirely by using a Hamiltonian formulation for the deterministic evolution of a quantum system \cite{Feynman:1981tf,Jordan:2011ci,Jordan:2011ne}.

There has been a significant effort to develop methods for simulating LGTs on quantum computers \cite{Banerjee:2012pg,Mueller:2022xbg,Ciavarella:2022zhe,Zohar:2012ay,Zohar:2012xf,Zohar:2013zla,Zohar:2014qma,Zohar:2015hwa,Zohar:2016iic,Klco:2019evd,Ciavarella:2021nmj,Bender:2018rdp,Liu:2020eoa,Hackett:2018cel,Alexandru:2019nsa,Yamamoto:2020eqi,Haase:2020kaj,Armon:2021uqr,PhysRevD.99.114507,Honda:2021ovk,Bazavov:2015kka,Zhang:2018ufj,Unmuth-Yockey:2018ugm,Unmuth-Yockey:2018xak,Kreshchuk:2020dla,Gustafson:2020vqg,multinode,Kreshchuk:2020aiq,Raychowdhury:2018osk,Raychowdhury:2019iki,Chakraborty:2020uhf,Wang:2022dpp,Davoudi:2020yln,Wiese:2014rla,Luo:2019vmi,Brower:2020huh,Mathis:2020fuo,Singh:2019jog,Singh:2019uwd,Buser:2020uzs,Bhattacharya:2020gpm,Barata:2020jtq,Kreshchuk:2020kcz,Ji:2020kjk,Bauer:2021gek,Gustafson:2021qbt,Hartung:2022hoz,Grabowska:2022uos,Murairi:2022zdg,Jahin:2022cws,Farrell:2022vyh,Li:2022ped,Farrell:2022wyt,Maxton:2022jgq,Asaduzzaman:2022bpi,2022arXiv220104546G,2022arXiv220812309G,2126556,Ciavarella:2021lel}. 
The benefits for high energy physics from quantum simulators stems from the ability to do real-time simulations and study finite density physics. However it is known that many time evolution methods introduce couplings to other energy modes which is undesirable \cite{carena2021lattice,2019arXiv191208854C,Gustafson:2020vqg}.
        
Implementations of the time evolution operator for quantum simulations involve approximating this operator \cite{Lloyd1073,Suzuki:1985,commeau2020variational,Bharti:2020cty,PhysRevX.7.021050,Lim:2021ivy,lau2022nisq,zoufal2021error,zagury2010unitary,barison2021efficient,yuan2019theory}. 
One method, Trotterization, break the Hamiltonian, $H$, is broken into commuting terms, e.g. potential ($V$) and kinetic ($K$), such that $e^{itH} \approx (e^{i\delta t/2K}e^{-i\delta t V}e^{i\delta t / 2 K})^{t/\delta t}$, which are easily implemented on a quantum computer \cite{Lloyd1073,Suzuki:1985,doi:10.1063/1.3078418}.
All approximations distort the Hamiltonian spectrum. Trotterization affects the spectrum by introducing terms proportional to $(t/n)^3$ multiplied by commutators such as $[K, [V,[V,K]]$; this significantly affects observables such as time dependent correlation functions \cite{carena2021lattice}.
The spectrum distortion induces couplings to other energy states \cite{2018PNAS..115.9456C,2019arXiv191208854C}.
Therefore one wants a band-pass filter that cuts out the energy modes unconnected to the states of interest; higher order Trotter product formulae do this \cite{Suzuki:1985}.
However, higher order Trotter products become increasingly more expensive in terms of gates and have diminishing accuracy returns \cite{2010JPhA...43f5203W}. 
The gate costs for quantum electrodynamics (QED) and quantum chromodynamics (QCD) simulations in Ref. \cite{kan2021lattice} is expensive; therefore, finding methods to bring these costs down is crucial. One method would be finding an alternative way to bring down the systematic errors from approximate time evolution (APE) other than improved operators.

% \begin{figure}[!th]
%     \centering
%     \includegraphics[width=\linewidth]{smear_test.pdf}
%     \caption{Demonstration of smearing for a representative time step $\delta t = 0.425$ using a second order Trotter product formula. The exact curve corresponds to the BCH mass oscillation.}
%     \label{fig:smearthisstuff}
%     \vskip-1em
\begin{figure}[!ht]
\vskip-2em
    \includegraphics[width=0.8\linewidth]{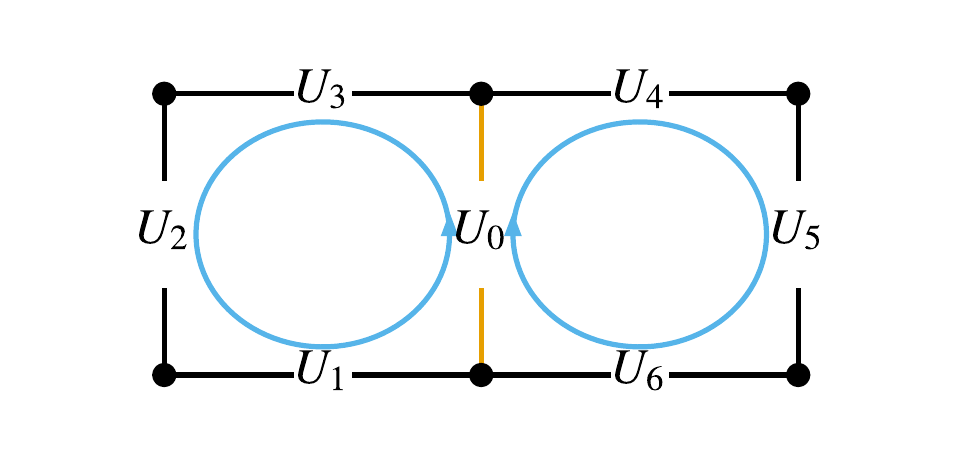}
    \vskip-2em
    \caption{Depiction of the plaquettes connected to a link in 2d. The orange line indicates the link to be smeared and the blue circles indicate the plaquettes that are taken as inputs to the smearing algorithm.}
    \label{fig:plaquettes}
\end{figure}

Classical LGT theory developed tools such as smearing \cite{PhysRevD.66.014501,joo2019pion,gusken1989non,PhysRevD.69.054501,2009PhRvD..80g4503Z,2009arXiv0901.2286B,Moran:2008ra,Basak:2005gi,GUSKEN1990361,ALBANESE1987163} and gradient flow \cite{Luscher:2009eq, Luscher:2011bx,2014PhRvD..89j5005B,2011JHEP...02..051L} as a method for dealing with high energy states in lattice configurations. This allows for classical simulations with $100\times-1000\times$ fewer statistics and coarser lattices \cite{Hasenfratz:2001hp,Durr:2004xu,ALBANESE1987163,phdthesisNikhil}.
All smearing methods averages neighboring fluctuations on a lattice configuration to remove ultraviolet (UV) contamination; gradient flow moves the configuration along the renormalization group and dampens high-momentum field modes which can be used as a continuous method of smearing \cite{Luscher:2011bx, Nogradi:2014aie,2014PhRvD..89j5005B,Fodor:2012td}. 
In addition smearing comes in two forms: operator smearing and link smearing. Operator smearing mitigates the excited state contamination from a lattice operator choices, link smearing suppresses the noise from UV fluctuations from the background field configurations themselves. 
Since one does not know the UV state explicitly one approximates them by the using the kinetic energy operator as a proxy. 
%Smearing smooths away the UV fluctuations on the underlying signal. 
%An example of real time smearing is shown in Fig. \ref{fig:smearthisstuff}.

Therefore one would like to have a quantum equivalent of classical lattice link smearing that is both cost effective and removes some couplings to high energy modes induced by approximate time evolution or state preparation. 
The quantum case of smearing can be understood as introducing a small imaginary component to the kinetic energy operator that will suppress their contribution to time dependent observables. 
In many cases these non-unitary algorithms can yield shorter local circuits than their unitary counter parts at the expense of some probability of failure \cite{2022APS..MARQ40002H, Lee:2019zze,choi:2020tio,Gustafson:2020vqg,2021PhRvA.104e2420H,Qian:2021wya}. 

In this context one can estimate the cost of smearing for a larger group that could approximate QCD, $S(1080)$. Using the chosen smearing strength is $\rho=0.2$ from Ref \cite{2022PhRvD.105k4508A} as an approximate value for the smearing parameter used for a quantum simulation of $S(1080)$, the quantum smearing operator has an approximate nonunitarity of $\eta=0.0005$ for a single link using the definition of $\eta$ in Ref. \cite{Zou:2022gbp}. If one estimates that a stochastic implementation such as in \cite{2021PhRvA.104e2420H} of the non-unitary operator succeeds with probability $1-\eta$, then every time all the links are smeared it will succeed with probability $(1-\eta)^{d*L^d}$ where $d$ is the number of dimensions and $L$ is the number of sites on the lattice in one direction. Using the volume as an estimate for gluon viscosity from Ref. \cite{kan2021lattice}, this would imply that for a $10^3$ lattice using smearing 50 times would succeed stochastically with probability $10^{-33}$. This is nearly impossible and the unitary method provided in this work would win out even with a subexponential increase in the number of required qubits.

In this letter, I develop a unitary quantum smearing algorithm based on classical stout smearing in a representation agnostic way. This quantum smearing algorithm scales linearly with the number of qubits and circuit depth compared to exponential costs with nonunitary operators. Using a discrete nonabelian gauge theory on a $2\times1$ lattice I demonstrate that the high energy modes are suppressed and the underlying physics is not distorted.

\textit{2 Stout smearing as a classical algorithm.}
%Smearing lattice configurations reduces the UV artifacts by replacing link variables $U$ by $U'$ such that the underlying configuration is smoother \cite{2009PhRvD..80g4503Z,2009arXiv0901.2286B,Moran:2008ra,Basak:2005gi,GUSKEN1990361,ALBANESE1987163}. 
%This happens by replacing a gauge link $U$ by $U'$ such that the underlying configuration is smoother. 
Stout smearing takes a linear combination of plaquettes connected to a link, see Fig. \ref{fig:plaquettes}, and then uses an exponential mapping of the linear combinations to transform the target link to a smeared link. The benefits of this smearing algorithm are that it smears out fluctuations at the lattice scale, preserves group structure, and is gauge invariant. Following the notation in Ref. \cite{Moran:2008ra}, I cover the basics of  stout smearing.

The staples connected to our target link $U_{\mu}(n)$ in the direction $\nu$ are defined as
\begin{equation}
    \label{eq:stapleop}
    \begin{split}
    C_{\mu,\nu}(n) = &U_{\nu}(n) U_{\mu}(n + \hat{\nu}) U^{\dagger}_{\nu}(n + \hat{\mu})\\
    & + U_{\nu}(n - \hat{\nu})^{\dagger}U_{\mu}(n - \hat{\nu})U_{\nu}(n - \hat{\nu} + \hat{\mu});
    \end{split}
\end{equation}
this is the sum of the plaquettes in Fig. \ref{fig:plaquettes}.
Next, one defines the linear combination of plaquettes as
\begin{equation}
    \label{eq:plaqs}
    \Omega_{\mu}(U_\mu(n)) = \Omega_{\mu}(n) =  \sum_{\nu\neq\mu}C_{\mu,\nu}(n) U^{\dagger}_{\mu}(n).
\end{equation}
A new variable, $\mathcal{Q}_{\mu}(n)$, defined as
\begin{equation}
\label{eq:Qop}
    \mathcal{Q}_{\mu}(n) = \frac{i}{2}\Big(\Omega_{\mu}(n) - \Omega^{\dagger}_{\mu}(n) - \frac{1}{N}Tr(\Omega_{\mu}(n) - \Omega^{\dagger}_{\mu}(n))\Big),
\end{equation}
creates a traceless Hermitian matrix which is a generator for a group element. 
The target link is then transformed to
\begin{equation}
    U'_{\mu}(n) = \mathcal{P} \lbrace e^{-i\rho\mathcal{Q}_{\mu}(n)}\rbrace U_{\mu}(n),
\end{equation}
where $\rho$ is a tunable parameter to determine how strong the smearing is and $\mathcal{P}$ indicates that some projection back onto the group may be required if the group is not sufficiently continuous. Hereinafter the element $\mathcal{P}\lbrace e^{-i\rho\mathcal{Q}(n)} \rbrace $ will be referred to as the shift element, $\mathcal{S}(\rho, \mathcal{Q})$. A graphical depiction of these terms is shown in Fig. \ref{fig:plaquettes}.

\textit{3 Stout smearing as a quantum algorithm.}
%Smearing does not work straightforwardly as a quantum algorithm because it irreversibly (nonunitarily) writes over information which is difficult to implement deterministically on a quantum computer. 
%While smearing works straightforwardly as a classical algorithm, it is a fundamentally a non-unitary transformation because it involves a irreversibly writing over information; this makes it difficult to implement on a quantum computer.
%One could implement this smearing procedure as a stochastic non-unitary operation however this comes with an exponential sampling cost \cite{2021PhRvA.104e2420H, 2022APS..MARQ40002H}.
A direct mapping of smearing encounters difficulties because it requires making a copy of the lattice; this is not allowed in quantum computation \cite{nielsen_chuang_2010,1970FoPh....1...23P,1982Natur.299..802W,1982PhLA...92..271D,Buzek:1996qw}. The copying procedure is required to avoid stroboscopic approximations. Therefore we need to alter the smearing algorithm to allow for reversibility which increases the required physical quantum resources compared to the classical algorithm.

Before introducing the algorithm, I will cover the necessary quantum operations needed for this algorithm. One requires three primitive group operations: $\mathfrak U_{\times}$ which multiplies two group elements together, $\mathfrak U_{-1}$ which inverts a group element, and $\mathcal{S}(\rho, \mathcal{Q})$ which generates the shift elements. The algorithm is summarized in Fig. \ref{fig:quantumsmear}.
 
\begin{figure}[h!t]
% \rule{\linewidth}[1pt]
\hrulefill
\begin{algorithmic}
\For{Link, $U_\mu(n)$, on the lattice}
\State Store each plaquette in Eq. (\ref{eq:plaqs}) onto scratch registers.
\State Generate the link variable corresponding to $\mathcal{S}(\rho, \mathcal{Q})$.
\State uncompute the plaquettes on the scratch register
\EndFor
\For{Link, $U_{\mu}(n)$, on the lattice}
\State Multiply the register with $\mathcal{S}(\rho, \mathcal{Q})$ to respective $U_{\mu}(n)$. 
\EndFor
\end{algorithmic}
\hrulefill
\caption{Pseudocode describing the quantum stout smearing algorithm.}
\label{fig:quantumsmear}
\end{figure}

%Implementations of multiply controlled not gates are also required which have fault-tolerant implementations \cite{Chuang:1996hw,PhysRevA.52.3457,2019arXiv190401671B}. This implies there is a transformation to a group element basis for any faithful representation of a gauge theory.  

%\paragraph*{Algorithm: Quantum Stout Smearing}
% \begin{enumerate}
%     \item Store each plaquette in Eq. (\ref{eq:plaqs}) onto a scratch register using $\mathfrak U_\times$ and $\mathfrak U_{-1}$. \vspace{-1em}
%     \item Generate the link variable corresponding to $\mathcal{P}\lbrace e^{i\rho\mathcal{Q}}\rbrace$ where $\mathcal{Q}$ is defined in Eq. (\ref{eq:Qop}). \vspace{-1em}
%     \item repeat steps 1 and 2 for every link on the lattice \vspace{-1em}
%     \item uncompute the plaquettes on the scratch registers \vspace{-1em}
%     \item multiply the links $\mathcal{P}\lbrace e^{i\rho\mathcal{Q}}\rbrace$ onto their corresponding physical lattice link \vspace{-1em}
% \end{enumerate}

\begin{figure*}
\includegraphics[width=0.85\textwidth]{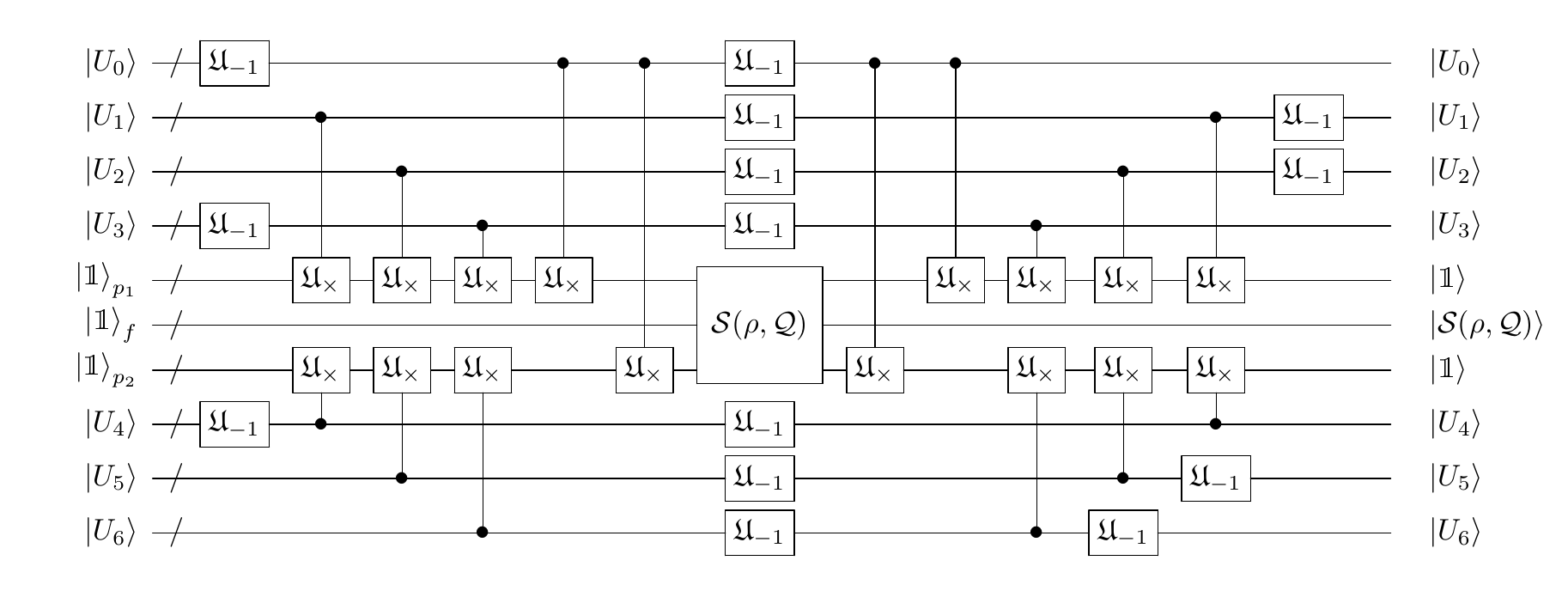}
\caption{Quantum circuit for constructing the projected element $\mathcal{S}(\rho,\mathcal{Q})$ on a 2d lattice. The forward slash indicates that the register maybe composed of multiple qubits. The registers $\ket{U_m}$ correspond to the links in Fig. \ref{fig:plaquettes}. The three register gate $\mathcal{S}(\rho,\mathcal{Q})$ takes the $\ket{}_{p_1}$ and $\ket{}_{p_2}$ scratch plaquette registers as inputs and outputs the closest group element to $e^{i\rho\mathcal{Q}}$ onto the scratch register $\ket{\mathbbm{1}}_f$.}
\label{fig:quantumstout}
\end{figure*}

This algorithm is representation agnostic; any Hamiltonian formulation requires a method to represent group element basis states. However, the process of implementing $\mathfrak{U}_{\times}$, $\mathfrak{U}_{-1}$, and $\mathcal{S}(\rho, \mathcal{Q})$ is representation dependent. 
Nevertheless, given group element basis exists this the function $\mathcal{S}(\rho, \mathcal{Q})$ is equivalent to a function that takes as inputs numbers and outputs a new number onto a clean scratch register which is a valid unitary operation on a quantum computer \cite{nielsen_chuang_2010}. A side effect of this algorithm is that a new lattice's worth of qubits is required every time the smearing operation is applied in order to ensure reversibility. Therefore there is a linear cost in the number of qubits to use this algorithm. 

We show in Fig. \ref{fig:quantumstout} how to construct the state $\ket{\mathcal{S}(\rho, \mathcal{Q})}$ on an ancilla register for any group and formulation. It is straight forward once all shift elements have been constructed to multiply these elements with their corresponding physical lattice link using the $\mathfrak U_{\times}$ operator. 

\begin{figure}[!ht]
\vskip-2em
\includegraphics[width=0.8\linewidth]{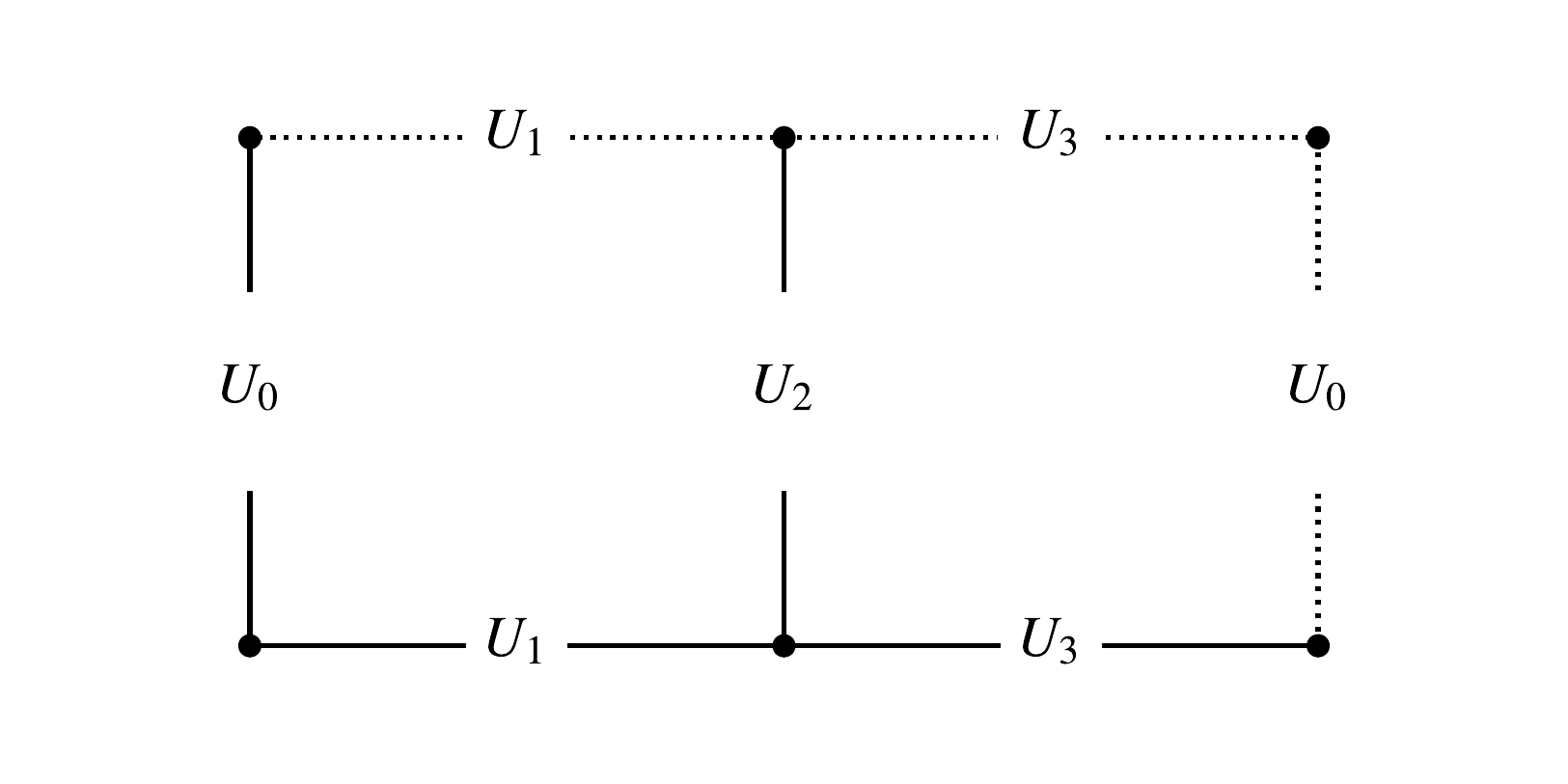}
\vskip-3em
\caption{2 Plaquette lattice for the $\mathbb{D}_4$ simulation. The dashed lines indicate repeated links from periodic boundary conditions.}
\label{fig:d4plaq}
    \centering
    \includegraphics[width=\linewidth]{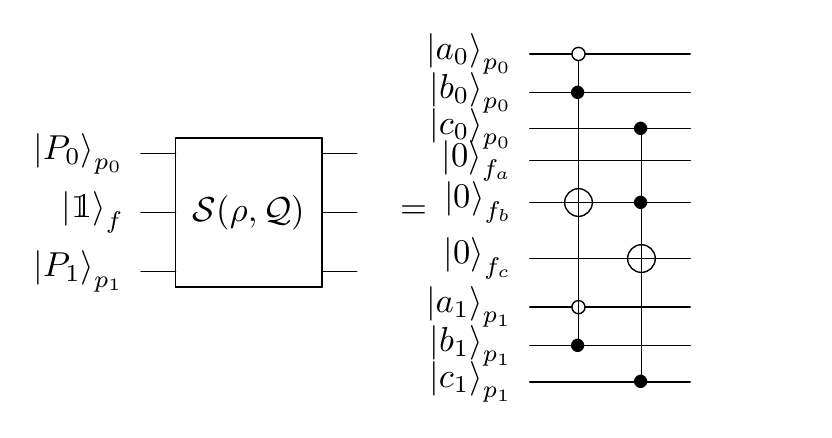}
    % \begin{equation*}
    %     \begin{gathered}
    %     \Qcircuit @R=0.8em @C=1em {
    %     \lstick{\ket{P_0}_{p_0}} & \multigate{2}{\mathcal{S}(\rho, \mathcal{Q})} & \qw \\
    %     \lstick{\ket{\mathbbm{1}}_f} & \ghost{\mathcal{S}(\rho, \mathcal{Q})}&\qw\\
    %     \lstick{\ket{P_1}_{p_1}} & \ghost{\mathcal{S}(\rho, \mathcal{Q})}&\qw
    %     }
    %     \end{gathered}
    %     ~~~
    %     =
    %     ~~~~~~~~
    %     \begin{gathered}
    %     \Qcircuit @R=0.8em @C=1em {
    %     \lstick{\ket{a_0}_{p_0}} & \ctrlo{7} & \qw & \qw \\
    %     \lstick{\ket{b_0}_{p_0}} & \ctrl{6} & \qw & \qw \\
    %     \lstick{\ket{c_0}_{p_0}} & \qw & \ctrl{6} & \qw \\
    %     \lstick{\ket{a_1}_{p_1}} & \ctrlo{4} & \qw & \qw \\
    %     \lstick{\ket{b_1}_{p_1}} & \ctrl{3} & \qw & \qw \\
    %     \lstick{\ket{c_1}_{p_1}} & \qw & \ctrl{3} & \qw \\
    %     \lstick{\ket{0}_{f_a}} & \qw & \qw  & \qw \\
    %     \lstick{\ket{0}_{f_b}} & \targ& \ctrl{1} & \qw  \\
    %     \lstick{\ket{0}_{f_c}} & \qw & \targ & \qw \\
    %     }
    % \end{gathered}
    % \end{equation*}
    \vskip-2em
    \caption{Quantum Circuit that implements the operator $\mathcal{S}(\rho=0.26,\mathcal{Q})$ for $\mathbbm{D}_4$.}
    \label{fig:Peipq}
\end{figure}

\textit{4 $\mathbbm{D}_4$ gauge theory example.}
It is illustrative to demonstrate smearing by a using discrete group which has the benefit of being defined explicitly in the group element basis.
A dihedral group, $\mathbbm{D}_4$, whose primitive gates have been derived in Refs. \cite{Lamm:2019bik, 2021arXiv210813305S}, is useful as a proof of principle example. 
The elements of the group are $g = (\sigma^x)^a (e^{i \pi / 2 \sigma^z})^{2b + c}$ with $0 \leq a, b, c \leq 1$ and $\sigma^x$ and $\sigma^z$ are two of the Pauli matrices. A group element $|g\rangle$ is repesented by the qubit state $|abc\rangle$.
This example uses a two plaquette theory with periodic boundary conditions as shown in Fig. \ref{fig:d4plaq} The Hamiltonian for this theory is
\begin{equation}
    \label{eq:ham}
    H = -\beta (\text{ReTr}(U_0 U_1 U_2^{\dagger} U_1^{\dagger} + U_2 U_3 U_0^{\dagger}U_3^{\dagger})) + \log(T_K),
\end{equation}
where $\log(T_K)$ is the kinetic term of the Hamiltonian, $K$, and is defined in \cite{Lamm:2019bik} and $-\beta(\text{ReTr}(U_0 U_1 U_2^{\dagger} U_1^{\dagger} + U_2 U_3 U_0^{\dagger}U_3^{\dagger}))$ is the potential term of the Hamiltonian, $V$. In this example $\beta = 0.75$.

An implementation of $\mathcal{S}(\rho, \mathcal{Q})$ for $\mathbbm{D}_4$ is necessary.
The realization of the circuit depends on the chosen value of $\rho$. In order to ensure that smearing occurs but is not too significant that it distorts the underlying physics, let $\rho=0.26$. For $\rho$ less than this value no smearing will take place.  Since $\mathbbm{D}_4$ is discrete, $\rho$ within given ranges will yield the same $\mathcal{S}(\rho, \mathcal{Q})$. Fig. \ref{fig:Peipq} shows the implementation of this quantum circuit for $\rho=0.26$. The time evolution operator $\mathbf{U}_n(t) = e^{-i t H}$, where $n$ is the Trotterization order, is at second order
\begin{equation}
    \label{eq:secondordertrot}
    \mathbf{U}_2(t; \delta t) = (e^{-i \delta t / 2 K} e^{-i \delta t V} e^{-i\delta t / 2 K})^{t / \delta t},
\end{equation}
and third order
\begin{equation}
    \label{eq:thirdordertrot}
    \begin{split}
    \mathbf{U}_3(t;\delta t) = & (e^{-i7\delta tV/24}e^{-i2\delta tK/3}e^{-i3\delta t V/4}e^{i2\delta t K/ 3}\\
    &e^{-i\delta t V/24}e^{-i\delta tK})^{t/\delta t}
    \end{split}
\end{equation}
where $K$ and $V$ are the kinetic and potential parts of the Hamiltonian. For these simulations $\delta t= 0.85$. This is an example which shows contributions from other eigenstates.

\begin{figure*}
\includegraphics[width=0.9\linewidth]{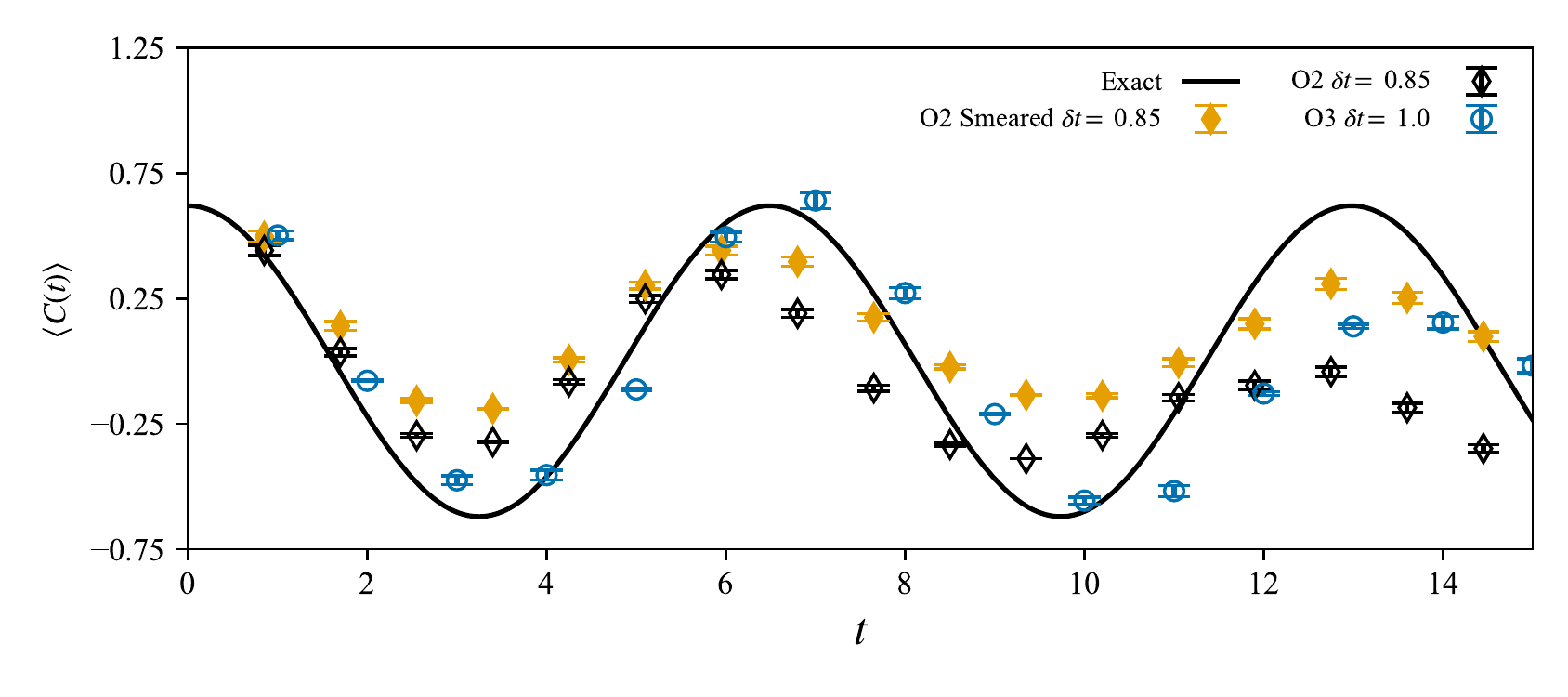}
\vskip-3em
\caption{Time evolution of the correlators $C(t)$, at $\beta = 0.85$ for second order Trotterization, O2, with and without smearing and third order Trotterization, O3, at $\delta t = 1.2$ for equivalent accuracy.}
\label{fig:smeartest}
\vskip-1em
\end{figure*} \fig{smeartest} shows the time evolution with $\delta t = 0.85$, of the plaquette correlator,
\begin{equation}
    \langle P(t)\rangle = \langle \Omega | U^{\dagger}(t; \delta t) P U(t;\delta t)|F\rangle,
\end{equation} where $\langle \Omega|$ is the gauge invariant ground state, $|F\rangle$ is the gauge invariant projection of the first excited state, and $P$ is the plaquette $\text{Re}\text{Tr}(U_0U_1U_2^{\dagger}U_1^{\dagger})$.  A third order Trotterization at $\delta t = 1.0$ which has nearly the same root mean square error as the smeared evolution is also shown. While the third order Trotterization is superior to the second order Trotterization with and with out smearing before $t = 10$, afterwards the higher energy states begin to distort the time evolution which is expected for a coarse Trotterization and aligns with the second order smeared Trotterization. The ancilla registers are reset in order to minimize the memory resource requirements on the classical simulations.

It is worth examining the resource costs of smearing in a fault tolerant perspective. Many error correcting codes for fault tolerant quantum computing have costly single qubit rotation operations such as the T-gate \cite{Chuang:1996hw,1996PhRvA..54.1098C,PhysRevLett.77.793,1996RSPSA.452.2551S,PhysRevA.54.4741,Eastin_2009}. For this reason T-gates are an important metric for algorithm costs on fault tolerant quantum computers.

\begin{figure}[ht]
\vskip-1em
\includegraphics[width=0.9\linewidth]{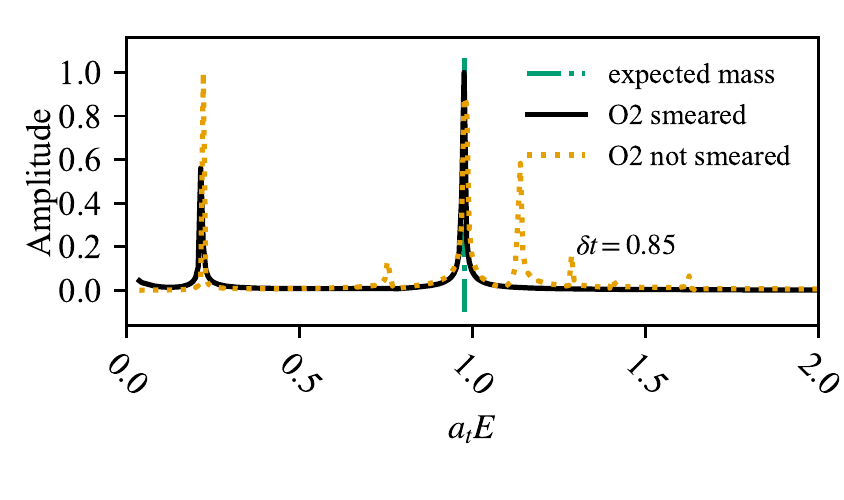}
\vskip-3em
\caption{Comparison of the energy spectrum with and without smearing at $\delta t = 0.85$ using second order Trotterization denoted O2.}
\label{fig:smearingfourier}
\end{figure}

\begin{table}[!ht]
\caption{T gate costs for various operations per Trotter step for the whole system on a $2\times1$ plaquette lattice for $\mathbb{D}_4$.}
\label{tab:tgatecost}
\begin{tabular}{cc}
\hline\hline
    operator & T gates \\\hline
    2$^{nd}$ order Trotter & 696 + 46$\log_2(1/\epsilon)$\\
    3$^{rd}$ order Trotter & 1008 + 131.1$\log_2(1/\epsilon)$\\
    smearing & 560 \\\hline\hline
\end{tabular}
\end{table}
The T-gate costs of a single second order Trotter step for $\mathbb{D}_4$, third order Trotter, and the smearing operator on the whole lattice in Tab. \ref{tab:tgatecost}. These costs are derived from the gates provided in Refs.  \cite{Lamm:2019bik,2021arXiv210813305S} and the single qubit gates are approximated using the repeat until success method with T gate cost $1.15\log_2(1/\epsilon)$ where $\epsilon$ is the desired gate infidelity \cite{2013arXiv1311.1074P}. 
The total cost for third order Trotterization for infidelity, $\epsilon=10^{-8}$ and $\delta t=1$ requires 1.75 times more T gates than a second order Trotterization at $\delta t = 0.85$ with smearing and 2.5 times as many T-gates if the third order Trotterization is used with $\delta t=0.85$. This T gate saving should increase for larger groups as the projection operation will not require approximations using T-gate synthesis.
The second order smeared evolution has fewer high energy oscillations than the unsmeared evolution. Information regarding energy states can be extracted from the fourier spectrum of the time series data. Examining the Fourier spectrum for the smeared and unsmeared evolution using second order Trotterization (see Fig. \ref{fig:smearingfourier}) shows that many high kinetic energy modes are mitigated. It is found that smearing does not uniformly suppress higher order energies.

% It is possible to estimate the cost of smearing for a larger group that could approximate quantum chromodynamics, $S(1080)$. Using \cite{2022PhRvD.105k4508A}, the chosen smearing strength is $\rho=0.2$, if we take this as an approximate value for the smearing parameter used for a quantum simulation of $S(1080)$ then we find that the quantum smearing operator has an approximate non-unitarity of $\eta=0.0005$ for a single link using the definition for the non-unitarity $\eta$ in Ref. \cite{Zou:2022gbp}. Then if we estimate that a stochastic implementation such as in \cite{2021PhRvA.104e2420H} of the non-unitary operator succeeds with probability $1-\eta$, we find that every time we want to smear all the links it will succeed with probability $(1-\eta)^{d*L}$ where $d$ is the number of dimensions and $L$ is the number of sites on the lattice in a given direction. Using the volume and lattice size as an estimate for gluon viscosity from Ref. \cite{kan2021lattice}, this would imply that for a $10^3$ lattice using smearing 50 times would succeed stochastically with probability $10^{-33}$. This is clearly impossible and the unitary method provided in this work would win out even with the linear increase in qubit cost. 

\textit{5. Conclusions}
This work developed a unitary algorithm for smearing real-time quantum simulations of LGTS. Somewhat analogously to classical LGT, this algorithm acts as expected by reducing higher energy modes. For the small lattice size investigated the benefits of smearing are noticable and achieves a factor of 2 reduction in T-gates compared to improved Trotterization. The computational costs scale linearly with the number of times smearing is applied compared to exponential costs for nonunitary evolution. The representation agnostic method in which this algorithm is presented allows it to be applied to a wide range of Hamiltonian formulations \cite{Raychowdhury_2020, Raychowdhury:2018osk,Dasgupta:2020itb,2022arXiv220607444M,2022arXiv220811789M,BROWER2004149, Beard_1998, Brower:2020huh,2022arXiv220812309G, Brower:1997ha,ciavarella2021trailhead,Zohar_2013,2019PhRvD.100a4506M, 2020PhRvD.102a4506M,Bazavov_2015,PhysRevLett.121.223201,Unmuth_Yockey_2018,2022PhRvD.105k4508A,2021arXiv210813305S,Ji_2020,2022arXiv220315541G,Alexandru_2019,2004PhRvL..92q7902V,2020npjQI...6...49S}%,2021PhRvA.103d2405D,2021arXiv210308056C}
for quantum simulation of LGTs and will likely bring down the cost of many fault tolerant applications. This opens the way to build smearing algorithms for larger groups that could approximate $SU(3)$ and $SU(2)$ as well as extend the method to the inclusion of dynamical fermions. 

\begin{acknowledgments}

I wish to thank Mike Wagman, Ruth Van de Water, Norman Tubman, Stuart Hadfield,  Henry Lamm, Judah Unmuth-Yockey, and Matthew Reagor for comments and advice.
This work is supported by the DOE QuantISED program through the theory  consortium ``Intersections of QIS and Theoretical Particle Physics" at Fermilab and by  the U.S. Department of Energy. Fermilab is operated by Fermi Research Alliance, LLC under contract number DE-AC02-07CH11359 with the United States Department of Energy.  
\end{acknowledgments}

% \nocite{*}

\bibliography{apssamp}% Produces the bibliography via BibTeX.

\end{document}